\documentclass[conference]{IEEEtran}
\usepackage{indentfirst}
\usepackage{graphicx}
\usepackage{framed}
\usepackage{pbox, flushend}
\usepackage{makecell}
\usepackage[hyperfootnotes=false]{hyperref}
\hypersetup{
 colorlinks=false,
 citecolor=black,
 linkcolor=black,
 urlcolor=black}

\usepackage{hyperref}
\usepackage{ifthen}
\usepackage{longtable}
\usepackage{graphicx}
\usepackage{array}
\usepackage{multirow}
\usepackage{booktabs}
 \usepackage{setspace}
\begin{document}

\title{An Investigation into Glomeruli  Detection\\ in Kidney H\&E and PAS Images using YOLO}

\author{
\IEEEauthorblockN{Kimia Hemmatirad$^1$, Morteza Babaie$^{1}$, Jeffrey Hodgin$^3$, Liron Pantanowitz$^3$, H.R.Tizhoosh$^{1,4}$}

\vspace{0.1in}
\IEEEauthorblockA{$^1$Kimia Lab, University of Waterloo, ON, Canada\\
$^3$ Department of Pathology, Michigan Medicine, University of Michigan, Ann Arbor, MI, USA\\
$^4$ Rhazes Lab, Artificial Intelligence and Informatics, Mayo Clinic, Rochester, MN, USA
}
}

\maketitle

\newpage
\begin{abstract}
Context -- Analyzing digital pathology images is necessary to draw diagnostic conclusions by investigating tissue patterns and cellular morphology. However, manual evaluation can be time-consuming, expensive, and prone to inter- and intra-observer variability. \\
Objective -- To assist pathologists using computerized solutions, automated tissue structure detection and segmentation must be proposed. Furthermore, generating pixel-level object annotations for histopathology images is expensive and time-consuming. As a result, detection models with bounding box labels may be a feasible solution.
Design -- This paper studies  YOLO- v4 (You-Only-Look-Once), a real-time object detector for microscopic images. YOLO uses a single neural network to predict several bounding boxes and class probabilities for objects of interest. YOLO can enhance detection performance by training on whole slide images. YOLO-v4 has been used in this paper  for glomeruli detection in human kidney images. Multiple experiments have been designed and conducted based on different training data of two public datasets and a private dataset from the University of Michigan for fine-tuning the model. The model was tested on the private dataset from the University of Michigan, serving as an external validation of two different stains, namely hematoxylin and eosin (H\&E) and periodic acid–Schiff (PAS). \\
Results -- Average specificity and sensitivity for all experiments, and comparison of existing segmentation methods on the same datasets are discussed.\\
Conclusions -- Automated glomeruli detection in human kidney images is possible using modern AI models. The design and validation for different stains still depends on variability of public multi-stain datasets. 
\end{abstract}

\IEEEpeerreviewmaketitle

\section{Introduction}
For investigations of tissue morphology and, as a result,  for making diagnostic conclusions, computational pathology approaches may offer fast and reliable solutions compared to conventional microscopy-based workflows. On the other hand, any manual evaluation of tissue samples can be time-consuming, costly, and subject to both inter- and intra-observer variability \cite{xing2016robust}. Consequently, researchers have recently focused their attention on automated solutions to detect and segment tissue structures in digital pathology whole slide images (WSIs). Many studies, such as determining tissue types, rely on the accuracy of tissue pattern segmentation, which is regarded as the foundation of automated image analysis. However, due to the complexity of tissue clustering into types, with architecture such as glands and organelles overlapping with each other, establishing precise segmentation is not a simple operation. This makes distinguishing these patterns from the tissue background, and mainly from each other a challenge. In addition, histopathological images may contain noise and artifacts created during image acquisition, as well as low contrast between foreground and background \cite{xing2016robust}. Segmentation models have been widely used in digital pathology to segment cells and other regions of interest \cite{irshad2013methods}. However, training these models needs pixel-level object annotations made by an expert. Detailed labels (pixel-level) for histopathology images are expensive, time-consuming, and hard to achieve \cite{khened2021generalized}. Moreover, in some of the applications      in histopathology, only detecting the position of the specific tissue pattern without precisely outlining  the  borders  may  be sufficient \cite{xing2016robust}. These techniques are called tissue pattern  detection, and are usually faster compare to segmentation methods. The main advantage of detection models  is  that  they construct a bounding box around the tissue of interest rather than pixel-level labelling, making its training much more convenient.

Deep object detectors typically consist of two parts: a backbone trained on ImageNet and a head used to forecast object classes and bounding boxes. One-stage object detectors and two-stage object detectors are the most common head types \cite{bochkovskiy2020yolov4}. Regions with convolutional neural networks (R- CNN) \cite{girshick2014rich} series, is a good example of a two-stage object detection category. YOLO (you-only-look-once) is one  of  the examples for one-stage object detectors \cite{bochkovskiy2020yolov4} which  has been studied and explored on two different applications for detecting specific tissue patterns in this paper.

YOLO is a simple concept with several advantages. To begin with, YOLO is very fast as it does not require a complicated pipeline for a regression problem. Furthermore, the mean average accuracy of YOLO is higher than that of comparable real-time systems. Therefore, the  network  can  be considered as a real-time object detector. Secondly, with YOLO, context information about object classes is encoded as well as their appearance during training and testing, unlike sliding window and region proposal-based approaches \cite{redmon2016you}. A popular object recognition approach, Fast R-CNN \cite{girshick2015fast}, may misidentify background patches in an image as objects. In comparison to Fast R-CNN, YOLO creates half the number of background errors \cite{redmon2016you}. And thirdly, YOLO learns to represent objects in a universally applicable way. YOLO surpasses the best detection algorithms like deformable parts models (DPM) and R-CNN when trained on natural images and evaluated    on art. YOLO is less likely to fail when applied to new domains or unexpected inputs because of its high degree of generalizability \cite{redmon2016you}.

In this paper, YOLO-v4 has been employed to find particular tissue patterns in WSIs. Comparisons on the same datasets, with segmentation approaches, will be performed. The histological evaluation of ``glomeruli'' is critical for identifying whether a kidney is transplantable \cite{marsh2018deep}. The Karpinski score, which includes the ratio of sclerotic glomeruli to total number of glomeruli in a kidney segment, is critical for determining the necessity for a single or dual kidney transplant \cite{altini2020deep}. Clinical symptoms, immunopathology, and morphological abnormalities are all factors that go into classifying glomerular disorders. To classify glomerular diseases, these anatomic structures need to be detected. Automated glomeruli identification frameworks for kidney biopsies conducted by pathologists can be quite helpful because manual examination of kidney samples is time-consuming and error-prone \cite{altini2020deep, marsh2018deep}. There are several segmentation methods to detect glomeruli  in kidney images \cite{bueno2020glomerulosclerosis}. However, these methods require pixel-level annotation of the images. In detection methods, only determining the location of a given tissue pattern, the glomerulus, is required without the need to precisely delineating its
borders.

In the field of histopathology, the lack of image data, annotation, and labels has always been a problem \cite{sudharshan2019multiple}. Hence, it is important to validate deep networks on their generalization capability. By training a network with public datasets, and then fine-tuning it with only limited  data  from a specific hospital or  specific  resource,  we  may  be  able to significantly improve the accuracy of the network on the validation set from the same resource.

In another application, YOLO-v4 as a detection network has been trained to recognize all glomeruli in a given kidney image. Multiple experiments were designed and carried out based on different training data from two public datasets to fine-tune the model, and tested on the private dataset from  the University of Michigan as an external validation on two differently stained tissues, namely periodic acid–Schiff (PAS) staining and hematoxylin and eosin (H\&E) staining.

\begin{figure*}[htb!]
\centering
%\captionsetup{justification=centering}
\includegraphics[width=\textwidth]{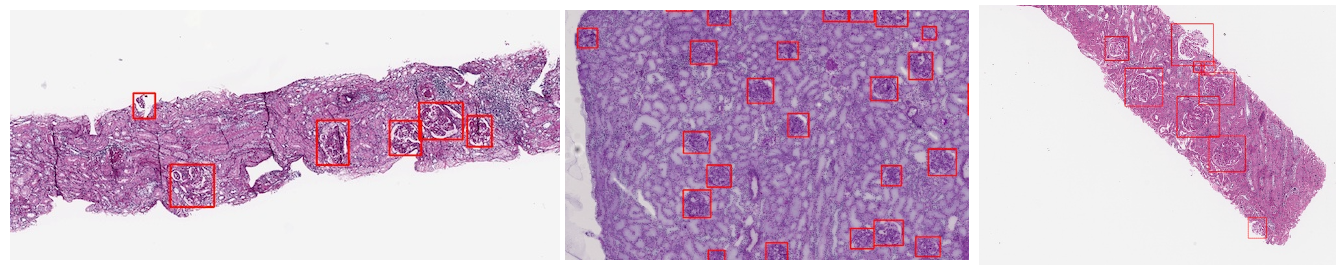}
\caption{Samples of the three training datasets: sample from the first public dataset AIDPATH (left), sample from the second public dataset HubMap (middle), and an example from the private dataset from the University of Michigan (right).}
\label{training_sample}
\end{figure*}

The first dataset is a public collection of  31 tiled TIFF (SVS)  WSIs. The annotation of the bounding boxes of these 31 WSIs has been performed by collaborating pathologists. This data is part of the WSI datasets generated within the European project AIDPATH  (source: http://aidpath.eu/). The second dataset has been used for the HubMap competition  (source: https://www.kaggle.com/c/hubmap-kidney-segmentation/overview). TIFF files ranging in size from 500MB to 5GB make up the dataset containing 8 WSIs for training, and 5 WSIs for subsequent testing. The segmentation annotation was provided for each of the WSIs in this competition. The generalization of the network has been tested by training on these two public datasets, followed by the external validation on the private dataset from University of Michigan. Another (private) dataset that has been used for training and fine-tuning the models has 7 PAS stained WSIs which has been collected from the University of Michigan annotated by an expert pathologist. In Figure \ref{training_sample} three samples of the training dataset for the network are shown.

The three datasets, two public and one private, have been used to design and conduct 14 experiments. These experiments have been trained on different combinations of public and private datasets. Results have been validated on the private dataset from the University of Michigan with two stains, 20 PAS stained WSIs and 16 H\&E stained WSIs. YOLO served as the detector network which will be described in details in the following sections.

\begin{figure}[htb!]
\centering
\includegraphics[width=\columnwidth]{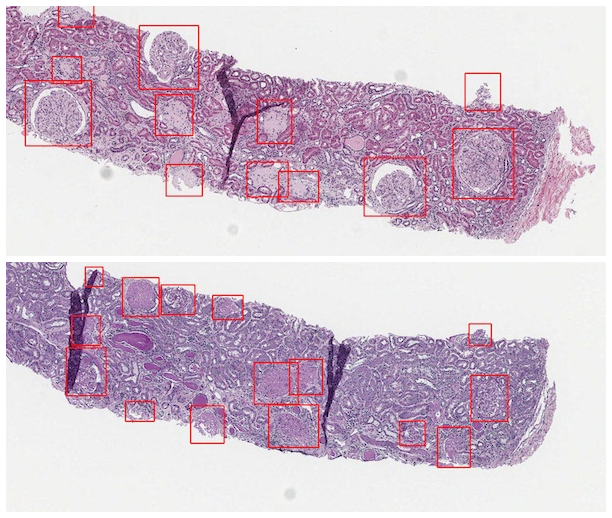}
\label{test_michigan}
\caption{Two samples of the private validation dataset along with annotated bounding boxes. On the top is a WSI   with H\&E-stained tissue, and the bottom shows a WSI with PAS-stained tissue.}
\end{figure}

In Figure \ref{test_michigan}, two samples of the private validation dataset along with the annotated bounding boxes have been shown. On the top is a  sample  of tissue  derived from a H\&E  stained  WSI,  and  on the bottom is a sample of tissue from a PAS stained WSI. The results, average specificity and sensitivity for all experiments. Comparison of existing segmentation methods on the same datasets are discussed in the results section. In general, one could observe that the average specificity and sensitivity are higher on the PAS validation set, because all of the images in the training dataset are PAS stained. Also, there is an improvement in average specificity and sensitivity while fine-tuning the network with only 7 PAS WSIs from the University of Michigan.

\section{Literature Review}
\begin{figure}[htb!]
\centering
\label{glomeruli-sample}
\includegraphics[angle=0,width=\columnwidth,keepaspectratio]{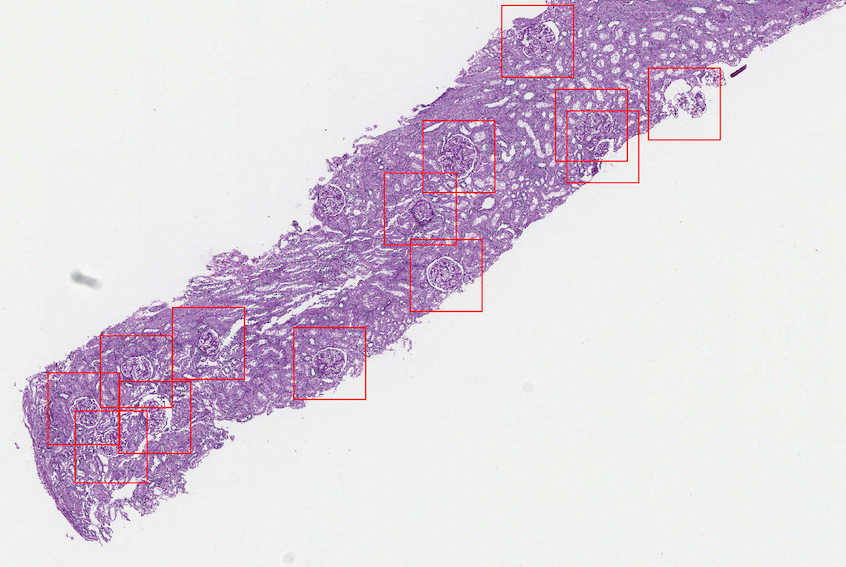}
\caption{Renal core biopsy showing annotated glomeruli.}
\end{figure}

A significant step in determining whether a kidney is transplantable is the histological examination of renal samples by experienced pathologists \cite{altini2020deep, marsh2018deep}. The histopathology evaluation of the number of globally sclerosed glomeruli in relation to the overall number of glomeruli is essential for accepting or rejecting a donor’s kidneys \cite{altini2020deep}. Mutliple glomerulocentric pathology classification systems are employed for native kidney diseases \cite{stokes2014morphologic,bajema2018revision, trimarchi2017oxford} emphasizing the central role of glomerular injury. In Figure \ref{glomeruli-sample} samples of glomeruli in kidney images are shown.

Waste and excess fluids are expelled from the human body by glomeruli, which are clusters of capillaries responsible for expulsion. It is possible to group glomerular disorders according to their clinical symptoms, etymology, immunopathology, or morphological changes \cite{altini2020deep, marsh2018deep}. A condition known as ``glomerulosclerosis'' is the result of the kidney lesion changing its morphology; this sclerosis can impact the kidney in many ways, depending on whether or not it is global or partial \cite{bueno2020glomerulosclerosis}. The number of glomeruli detected in each kidney biopsy should be counted in daily practice. Per kidney biopsy, about 20 to 30 cuts are made \cite{bueno2020glomerulosclerosis}. Additionally, glomeruli that are completely sclerosed must be noted (the entire glomerulus). Detection of localized sclerosis will provide further information regarding the patient's condition. Each pathology report should include this information because the number of glomeruli assessed must be representative enough to determine a diagnosis \cite{vickers2017animal}. On the other hand, if the sample has numerous sclerosed glomeruli, this may suggest that the patient has chronic kidney disease with dead glomeruli. As a result, the patient may not be suited for some medications, which will help to define adequate treatment \cite{vickers2017animal}. This information is also entered into the national register for glomerulonephritis  \cite{bueno2020glomerulosclerosis}. Time-consuming and tiresome, the count of glomeruli is a painstaking process. Because of this, image processing methods that can identify and categorize the glomerulus are needed. 

With the emergence of deep learning networks, various options for computer vision tasks such as glomeruli object identification, semantic segmentation, and instance segmentation became available \cite{bueno2020glomerulosclerosis}. For instance, some works provide a detailed assessment of object identification and instance segmentation algorithms \cite{zhao2019object}. Others provide a complete review of semantic segmentation \cite{garcia2017review}. Several recent research efforts in digital pathology have used deep neural networks for glomeruli detection and segmentation \cite{ledbetter2017prediction, kawazoe2018faster, cascarano2019innovative, altini2020semantic, gallego2018glomerulus, marsh2018deep, kato2015segmental, simon2018multi, temerinac2017detection, bueno2020glomerulosclerosis}.

For glomeruli detection, YOLO has been applied on kidney images for the first time in this  paper  and  compared with the existing segmentation method U-Net, using the same validation dataset. There are two different tissue stains in the validation dataset.

\subsection{Tissue Staining}
Staining is used to emphasize essential characteristics of the tissue, as well as improve contrast. Hematoxylin is a common stain dye used in this technique that gives the nuclei a bluish hue, whereas eosin (another stain dye used in histology) gives the cell’s cytoplasm a pinkish tint \cite{alturkistani2016histological}.

\subsubsection{Periodic Acid-Schiff (PAS)} 
A staining technique called PAS is used in histochemistry to show that carbohydrates and carbohydrate compounds like polysaccharides, mucin, glycogen, and fungal cell wall components are found in cells. PAS has been used to look for glycogen in places like the skeletal muscle, liver, and heart muscle. PAS staining works with both formalin-fixed, paraffin-embedded (FFPE), and frozen tissue sections \cite{wittekind2003traditional}. In renal pathology PAS stain is particularly useful to highlight basement membranes. 

\subsubsection{Hematoxylin and Eosin (H\&E) Staining} 
There are two types of histological stains that come together to make H\&E: Hematoxylin and Eosin. The hematoxylin stains cell nuclei  purple, and eosin stains the extracellular matrix as well as     the cytoplasm pink. Other anatomic tissue structures take on different shades and hues of  these  two  colors  \cite{chan2014wonderful}.  There are two parts of a cell including the nucleus and the cytoplasm. Pathologists can easily tell them apart, and the overall patterns of coloration from the stain show the general layout and distribution of cells and give an overall impression tissue morphology \cite{wittekind2003traditional}.

\section{Materials \& Methods}
\label{Chap:3}
\subsection{Method}
Predicting one or more object locations, determining their classes, and drawing a bounding box  around  the  object  is the definition of an object detection task. In many existing detection systems, multiple classifiers are applied to an image at many locations and scales to calculate the high-scoring regions of the image for detecting a region of interest. In this paper, the YOLO approach (You Only Look Once) \cite{redmon2016you} has been trained for both applications on detecting tissue patterns in WSIs; one is artifacts and manual ink-markers detection, and the second is glomeruli detection in kidney images.

One of the essential advantages of YOLO over classifier-based systems is the speed of this model.  YOLO  is faster than R-CNN with greater than 1000 times performance \cite{girshick2014rich}, and 100 times faster than Fast R-CNN \cite{girshick2015fast}. Predictions based on YOLO are with a single network evaluation, while R-CNN requires many network evaluations for a single image. More importantly, as an object detector, YOLO does not require detailed pixel-level annotation; labels for YOLO are just bounding boxes around the target objects.

\subsubsection{Network Architecture of YOLO}
 By combining separate components of other object detection networks, like the ones using a sliding window, or region-based techniques, YOLO can predict all image objects for all the classes based on the information from the whole image only by looking at the image once \cite{redmon2016you}. In other words, the network models the entire image at once along with all of its individual objects. End-to-end training and real-time speeds are made possible by the YOLO architecture while high average accuracy is maintained \cite{redmon2016you}.
An $S\times S$ grid is generated on any given image. A grid cell is responsible for identifying an object whose center lies within that grid cell. Boxes and confidence ratings are predicted for each grid square. If the model is certain that the box contains an object, it will give it a high confidence score \cite{redmon2016you}. This confidence score is calculated based on

\begin{equation}
Pr(Object) = IOU^{truth}_{pred} 
\end{equation}

The confidence score should be 0 if there is no predicted object present in a cell. If there is at least one predicted object in that cell, for the confidence score to be accurate, it must   be equal to the intersection over union (IOU) between the predicted box and the ground truth. The probabilities of each $C$ conditional class,  $Pr(Class_{i} | Object)$,  are  also  predicted in each grid cell. The location of object's grid cell determines these probabilities. No matter how many boxes $B$  there are   in a grid cell, the network can only forecast one set of class  probabilities. For the evaluation of the network, the network of computes the class-specific confidence scores for each box (each class $C_i$ and object $O$) based on

\begin{equation}
    Pr(C_{i} | O) * Pr(O) * IOU^{truth}_{pred} = Pr(C_i) * IOU^{truth}_{pred}
\end{equation}

Both the likelihood that a certain class will be found in the box and how well the predicted box will fit the item are represented by these scores \cite{redmon2016you}.

\begin{figure*}[h]
\centering
%\captionsetup{justification=centering}
\includegraphics[angle=0,width=0.9\textwidth,keepaspectratio]{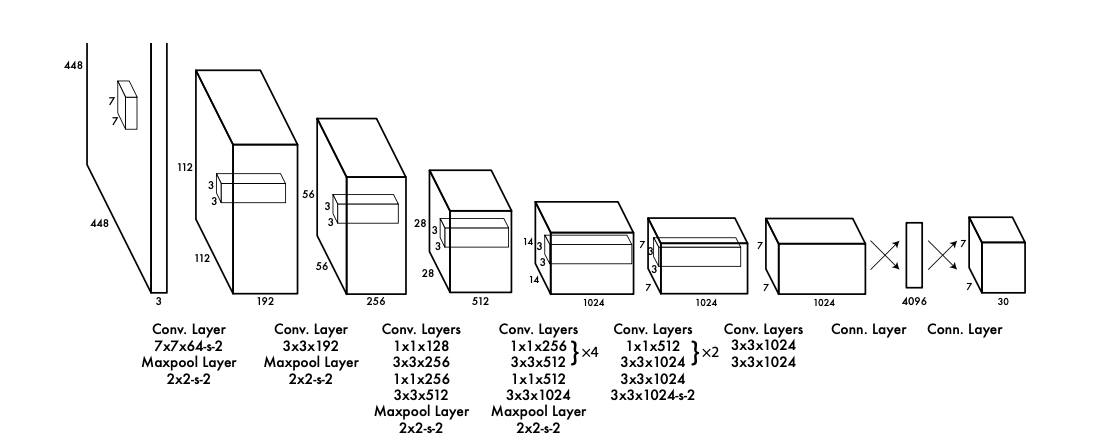}
\caption{YOLO Architecture: 24 convolutional layers and two fully linked layers make up the detecting network. This  image has been reproduced from \cite{redmon2016you}.}
\label{yolo-architecture}
\end{figure*}

YOLO was inspired by GoogleNet model for image classification \cite{szegedy2014going}. It has 24 convolutional layers, followed by two fully connected layers make up the detecting network. The full network has been shown in Figure \ref{yolo-architecture}.

\subsubsection{YOLO-v4}
The YOLO-v4 \cite{bochkovskiy2020yolov4} model has been trained  for both applications in this paper. The codes of all steps involved are available on GitHub (source: https://github.com/AlexeyAB/darknet), and in this experiment,  the codes have been modified in a way to train a custom dataset. The implementation of a new architecture in the backbone in YOLO-V4 compared to YOLO-V3 has made an essential improvement in the mAP (mean Average Precision) and the number of FPS (Frame per Second) by 10\% and 12\%, respectively, when trained and tested on COCO dataset (source: https://cocodataset.org/). The new architecture in the backbone is a deep neural network composed mainly of convolution layers, and the main objective is to extract features. The backbone selection is a key step and can improve object detection performance; mostly pre-trained neural networks are used to train the backbone \cite{bochkovskiy2020yolov4}.

\subsection{Dataset}
Kimia Lab and the pathology department of the University of Michigan are collaborating on a project for developing a computational kidney disease diagnosis model. As a part of this project, Kimia Lab has received a glomeruli dataset with bounding box annotations created by nephropathologists. To expand the training data two different public datasets, plus     a private dataset from the University of Michigan have been used in this study.

\begin{figure}[htb!]
\centering
\includegraphics[angle=0,width=\columnwidth]{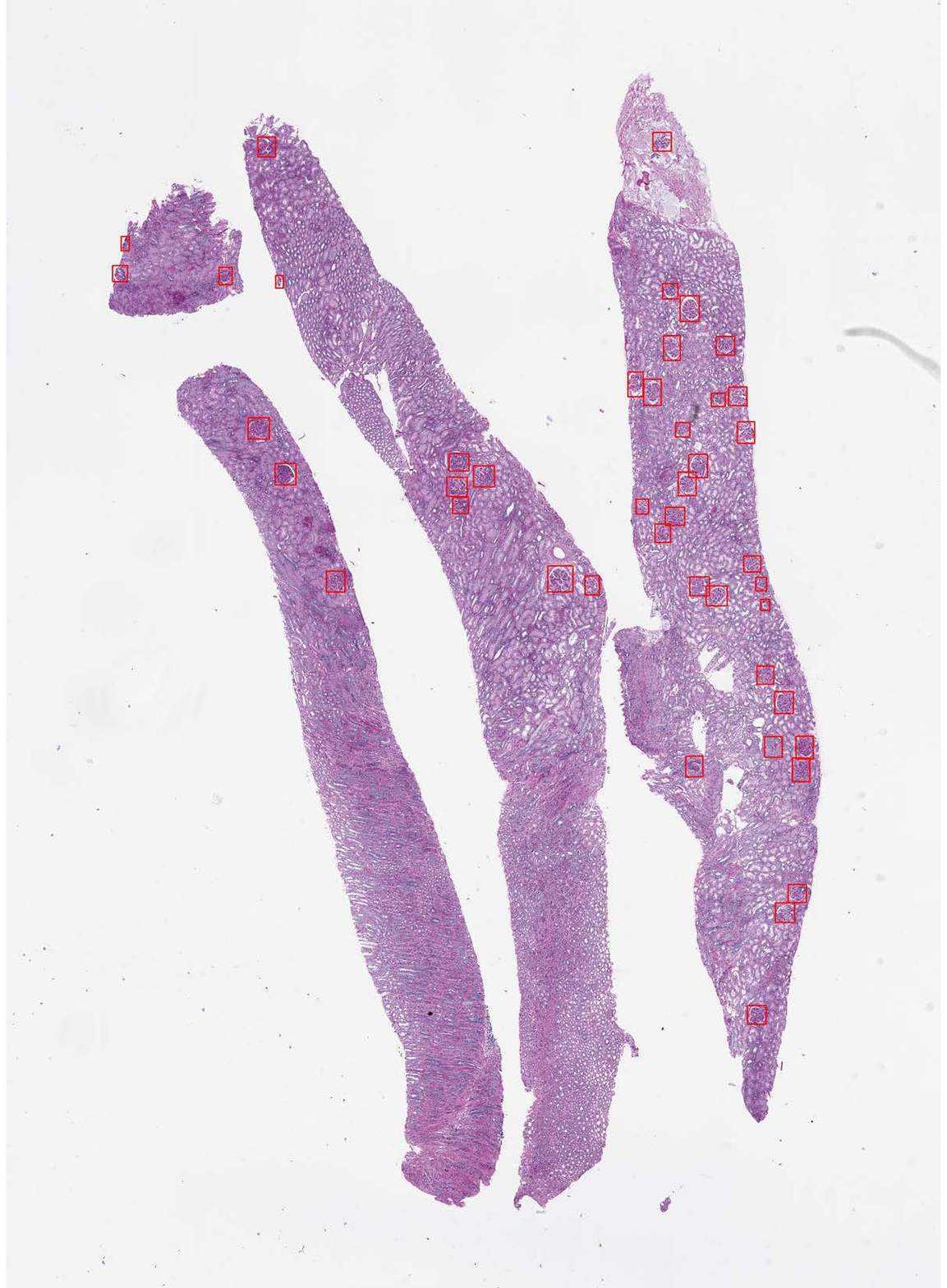}
\caption{Annotated WSI sample from public dataset 1.}
\label{public_1_sample}
\end{figure}

\paragraph{Public Dataset 1} 
The  first  public  dataset  consists of 31 WSIs in SVS format. With the size range  between $21651 \times 10498$ pixels and $49799 \times 32359$ pixels acquired at 20x to preserve image quality and information while requiring significantly less computational time than images taken at other magnifications \cite{bueno2020data}. A glomerulus may lose structural information due to the lower resolution and poor image quality. It is also important to note that employing magnifications such as 40x would increase the model size, slowing down     the training process \cite{bueno2020glomerulosclerosis}. This data is part of the WSI datasets generated within the \emph{European project AIDPATH}  (source: http://aidpath.eu/). A biopsy needle with an outside diameter of between 100 nm and 300 nm was used to obtain tissue samples. Once the paraffin blocks were ready, the tissue portions were cut into 4 um pieces and coloured with through PAS staining \cite{bueno2020data}. It is common to employ PAS stain to color polysaccharides found in kidney tissue and to highlight glomerular basement membranes be- cause of its effectiveness \cite{robinson2012renal}. These images contain different types of glomeruli labeled by Bueno et al. approach \cite{bueno2020glomerulosclerosis}. This dataset has two parts, DATASET A, which contains the raw  31 WSIs, and DATASET B, which is 2340 glomeruli images, 1170 normal glomeruli and 1170 sclerosed glomeruli. Because of the lack of exact coordinates of the extracted glomeruli,  the exact coordinates of the glomeruli bounding boxes were extracted by a pathologist.% at Kimia Lab %\footnote{Dr. Ricardo Gonzalez, an anatomic pathologist, visiting Kimia Lab from May 2021 till April 2022.}.
An annotated WSI sample of the first public dataset has been shown in Figure \ref{public_1_sample}.

\begin{figure}[h!]
\centering
\includegraphics[angle=0,width=0.9\columnwidth]{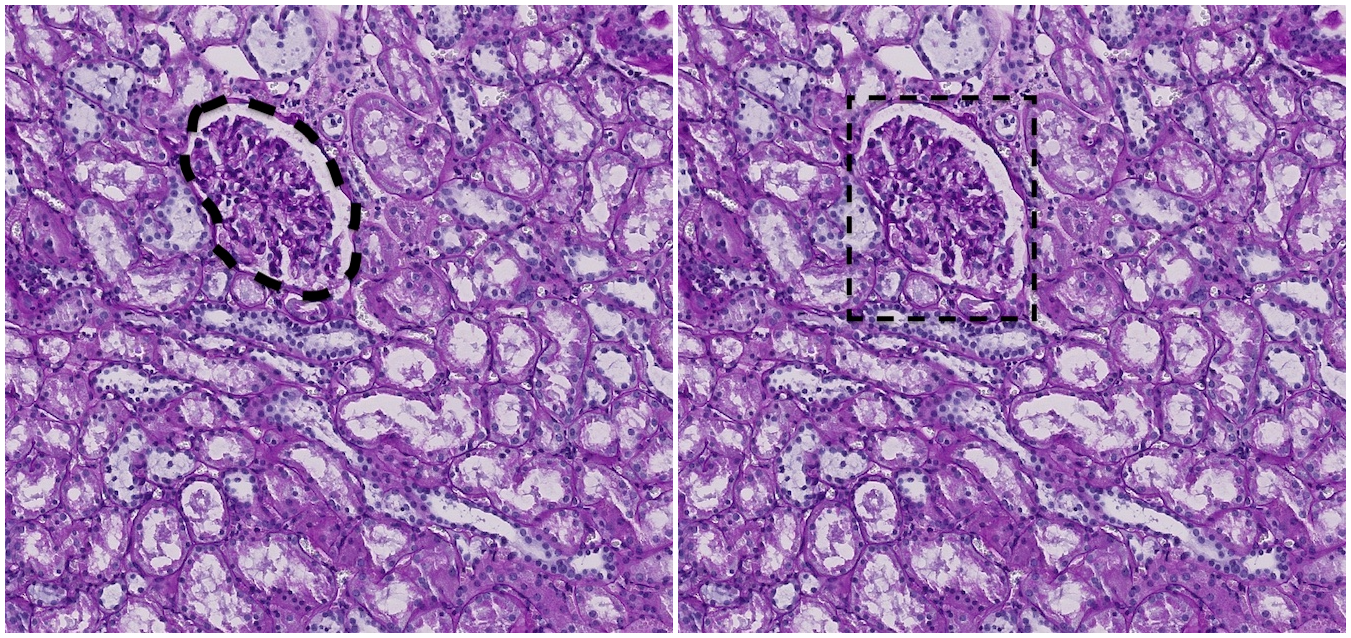}
%\captionsetup{justification=centering}
\centering
\caption{Extracted bounding boxes (right) from manual delineations of a glomerulus (left).}
\label{convert-coordinates}
\end{figure}

\begin{figure}[htb!]
\centering
\includegraphics[angle=0,width=\columnwidth]{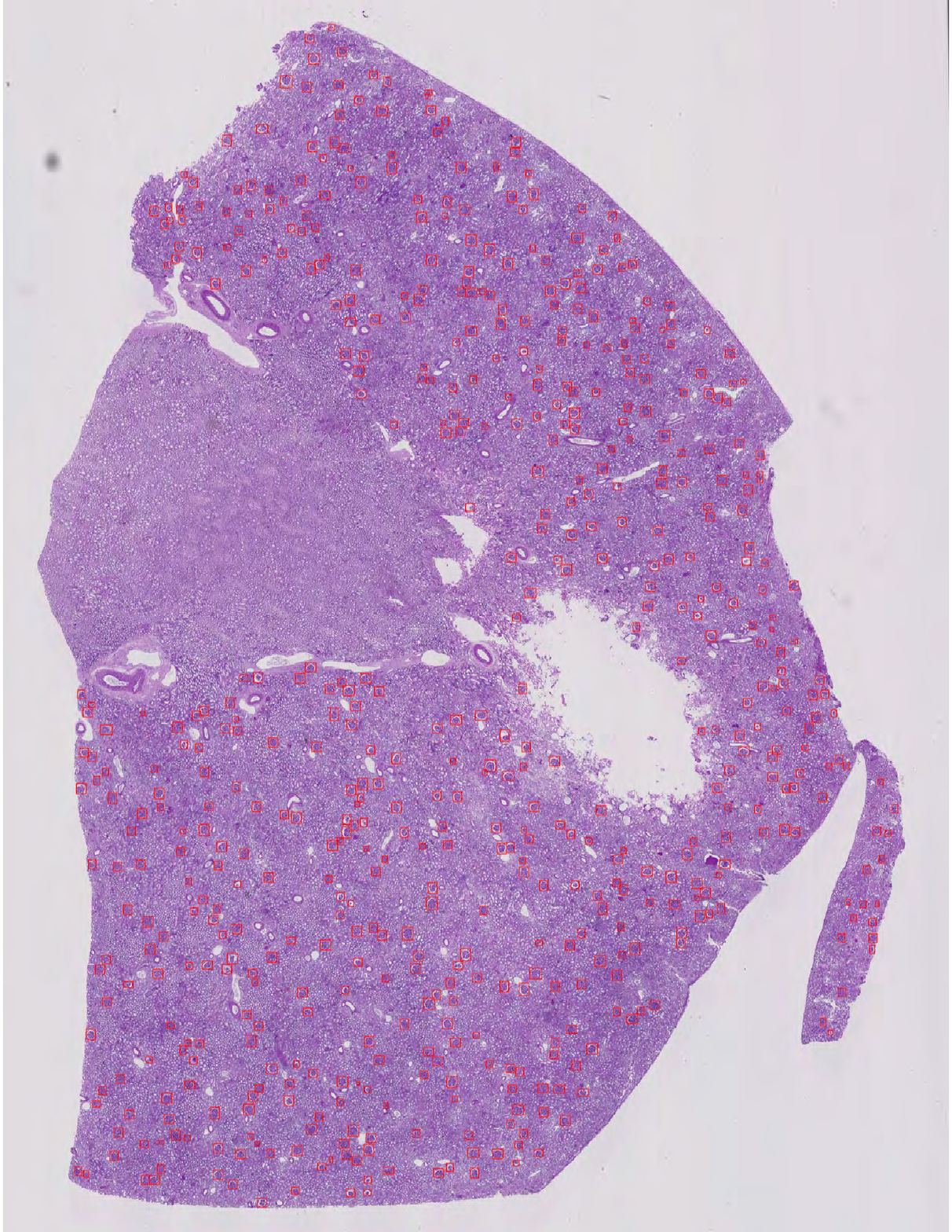}
\caption{Annotated WSI sample from public dataset 2.}
\label{public_2_sample}
\end{figure}

\paragraph{Public Dataset 2}
This dataset has been used for \emph{HubMap competition}   (source: https://www.kaggle.com/c/hubmap-kidney-segmentation/overview). TIFF files ranging in size from 500MB to 5GB make up the dataset containing 8 images for the training and 5 images for the test. RLE-coded and uncoded (JSON) annotations are included in the training set. The annotations identify glomeruli that have been divided into sections. Also, anatomical structural segmentations are included in both the training and public test sets. The bounding  boxes  of  these anatomical structures for using these annotations for the YOLO object detector have been created based on manual contours. Figure \ref{convert-coordinates} is an example of the procedure to generate a bounding box from manual delineation. This bounding box is found by calculating the upper left most and lower right- most coordinates in the delineation. An annotated WSI of the second dataset has been shown in Figure \ref{public_2_sample}.

\begin{figure}[htb!]
\centering
\includegraphics[angle=0,width=\columnwidth]{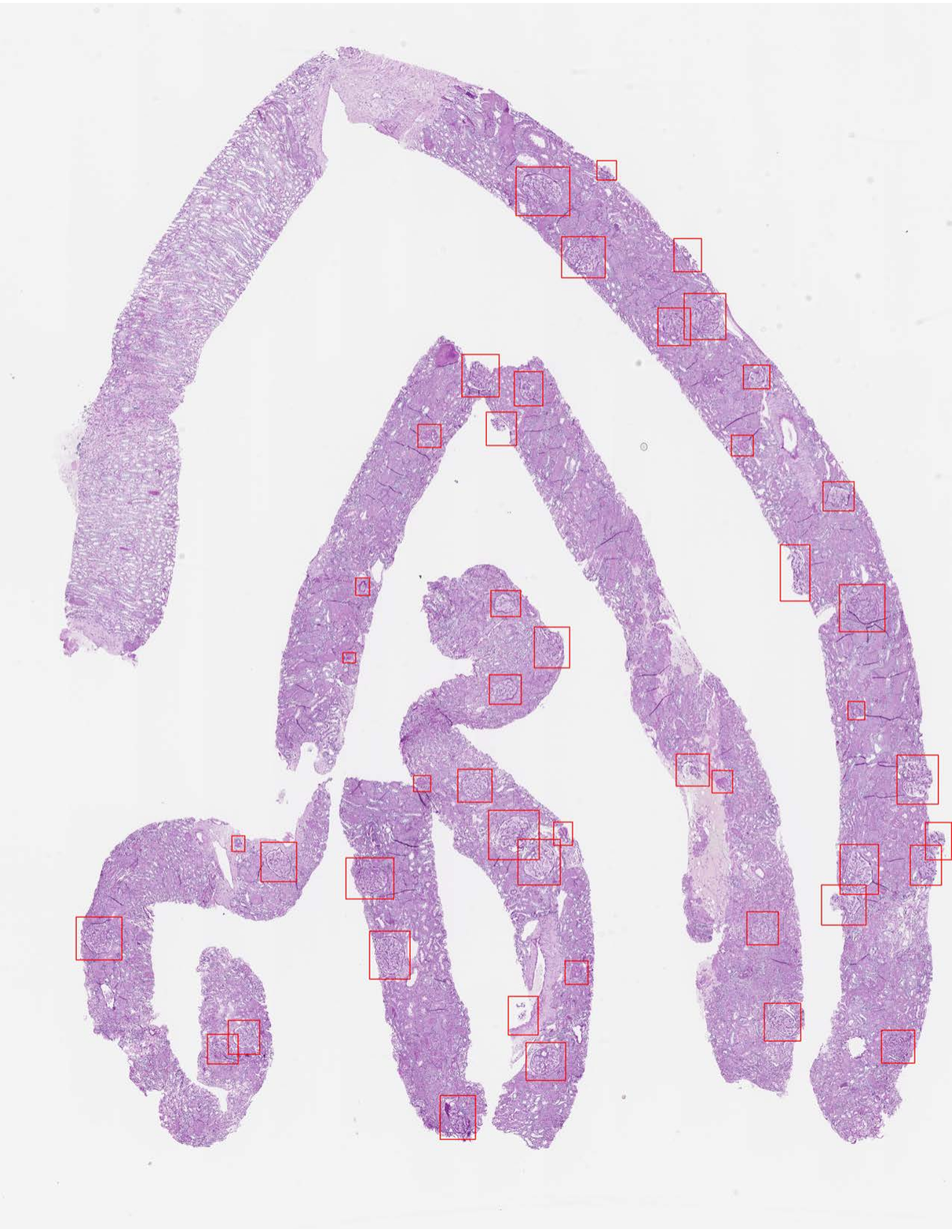}
\caption{Annotated WSI sample from the University of Michigan's private dataset.}
\label{sample-michigan-wsi}
\end{figure}

\paragraph{University of Michigan Data}
This private dataset has been collected from the University of Michigan and annotated by an expert pathologist. The training dataset consists of 7 Periodic acid–Schiff (PAS) stained WSIs for  fine-tuning the models that have been trained on the mentioned public datasets. Annotated WSI sample of this dataset has been shown in Figure \ref{sample-michigan-wsi}. Beside these 7 PAS stained WSIs for training,   20 PAS stained WSIs, and 16 H\&E stained WSIs have been used for validation. The images show that the University of Michigan data has the same type as the first public dataset, namely needle biopsy images, unlike the second public dataset, which are surgical excisions. This difference would affect the results obtained by each public dataset.

\subsection{Experiments}
A total of 7 different combination of datasets (using two public datasets, and a private dataset) selected for the training of YOLO object detector, resulting in a total of 7 different models. The 7 training datasets have been evaluated on two different validation datasets with different stains from the University of Michigan: One contains 20 PAS stained images and the other one contains 16 H\&E stained images. All experiments, along with the explanation of the training and  the validation dataset, have been reported in Table  \ref{glomeruli-experiments}. And   the configurations of the network for all 7 different training datasets have been described in Table \ref{config-network-2}. In this table,
\begin{itemize}
    \item \textbf{Batch} stands for how many images are used in the for- ward pass to compute a gradient and update the weights via back-propagation,
    \item \textbf{Subdivisions} stands for the number of blocks in which the batch is subdivided,
    \item \textbf{Policy} means using the steps and scales parameters bellow to adjust the learning rate during training,
    \item \textbf{Steps} means adjust the learning rate after 3200 and 3600 batches,
    \item \textbf{Scales} means re-scale the current learning rate by the corresponding factor once the number of steps is reached,
    \item \textbf{Max batches} is the maximum number of iterations,
    \item \textbf{Filters} stands for how many convolutional kernels there are in a layer, and 
    \item \textbf{Activation} defines the activation function.
\end{itemize}

\begin{table*}[!p]
%\captionsetup{justification=centering}
\centering
\caption{The configuration of the network for all training datasets for glomeruli detection}
\label{config-network-2}
\resizebox{\linewidth}{!}{%
\begin{tabular}{@{}c|c|c|c|c|c|c|c|c@{}}
learning rate & batch & subdivisions & policy & steps      & scales  & max batches & filters & activation \\ \midrule
0.001         & 40    & 16           & steps  & 4800, 5400 & 0.1,0.1 & 6000        & 18      & linear    
\end{tabular}%
}
\end{table*}

\begin{table*}[h]
\begin{center}
%\captionsetup{justification=centering}
\centering
\vspace{0.2in}
\caption{All 7 training sets along with the test experiments using public datasets 1 and, the private dataset from University of Michigan.}
\label{glomeruli-experiments}
\resizebox{\linewidth}{!}{\begin{tabular}{|c|| >{\centering\arraybackslash}m{7cm}| >{\centering\arraybackslash}m{7cm}|} 
\hline
\centering
    Experiment & Training Dataset & Test Dataset \\ \hline \hline
    
        1  & 31 WSIs from public dataset 1 & 20 PAS WSIs from UMICH dataset \\ \cline{3-3} & & 16 H\&E WSIs from UMICH dataset \\ \hline \hline
    
        2  & 31 WSIs from public dataset 1, fine-tuned with 7 PAS  WSIs from UMICH dataset & 20 PAS  WSIs from UMICH dataset \\ \cline{3-3} & & 16 H\&E  WSIs from UMICH dataset \\ \hline \hline
    
        3  & 8 WSIs from public dataset 2 & 20 PAS  WSIs from UMICH dataset \\ \cline{3-3} & & 16 H\&E  WSIs from UMICH dataset \\ \hline \hline
    
        4  & 8 WSIs from public dataset 2, fine-tuned with 7 PAS  WSIs from UMICH dataset & 20 PAS  WSIs from UMICH dataset \\ \cline{3-3} & & 16 H\&E  WSIs from UMICH dataset \\ \hline \hline
    
        5  & 31 WSIs from public dataset 1, and 8 WSIs from public dataset 2 & 20 PAS  WSIs from UMICH dataset \\ \cline{3-3} & & 16 H\&E  WSIs from UMICH dataset \\ \hline \hline
    
        6  & 31 WSIs from public dataset 1, and 8 WSIs from public dataset 2, fine-tuned with 7 PAS  WSIs from UMICH dataset & 20 PAS  WSIs from UMICH dataset \\ \cline{3-3} & & 16 H\&E  WSIs from UMICH dataset \\ \hline \hline
    
        7  & 7 PAS stained WSIs from UMICH dataset & 20 PAS  WSIs from UMICH dataset \\ \cline{3-3} & & 16 H\&E  WSIs from UMICH dataset \\ \hline 
\end{tabular}}
\end{center}
\end{table*}

Many studies have been performed to identify glomeruli functional tissue units in human kidneys. Recently, there was  a Kaggle competition, \emph{Hacking the Kidney}, launching to segment glomeruli in kidney images  (source: https://www.kaggle.com/competitions/hubmap-kidney-segmentation). The dataset provided for the competition was public dataset 2, discussed in the dataset section.  TIFF  files,  ranging  in  size  from  500MB  to 5GB, make up the dataset  containing  eight  images  for  the training and five images for the test. RLE-coded and uncoded (JSON) annotations are included in the training and validation sets. The authors of a study \cite{godwin2021robust} compare the five winning algorithms between more than a thousand teams that participated in the above competition. They assess the accuracy and performance of the five top algorithms, and the codes are available online (source: https://github.com/cns-iu/ccf-research-kaggle-2021/). To compare a segmentation model with the detection model in this paper, the first team’s algorithm has been chosen as  the benchmark. The accuracy on the same validation dataset
i.e. 20 PAS stained images and 16 H\&E images from the University of Michigan has been calculated based on the explanation for the winning proposal  (source: https://www.kaggle.com/c/hubmap-kidney-segmentation/discussion/238198). They have used a single U-Net SeResNext101 architecture with Convolutional Block Attention Module (CBAM), hypercolumns, and deep supervision. Their network read 1024$\times$  1024 pixel patches  and then downsample them to 320$\times$ 320 patches. SGD is the optimizer for their model, trained using binary cross-entropy. 

  Training is performed for  20  epochs,  with  a  learning  rate of $10^{-4}$ to $10^{-6}$ and a batch size of 8 images. Their final weights trained on the whole training dataset have been used to validate and compare their network on the University of Michigan dataset, which contains 20 PAS stained images and 16 H\&E stained images. The results are provided in section \ref{Chap:4}. Note that this is not possible to fine-tune the mentioned segmentation model with the external validation set (University of Michigan WSIs) as the external WSIs do not contain the pixel-level annotation. For comparing the segmentation model with YOLO, the segmentation area is enclosed with the smallest possible rectangle (the upper left most and lower right-most coordinates) and use these rectangles as the segmentation model output. Figure \ref{convert-coordinates} depicts the process visually.

\section{Experiments \& Results}
\label{Chap:4}

Immunopathology, clinical symptoms, and morphological abnormalities are all factors that go into classifying glomeruli disorders \cite{mastrangelo2020clinical}. To classify the glomeruli diseases, these objects need to be detected first. Therefore, the average sensitivity and  specificity of the detection matters. 

By having True Positives as $TP$, False Positives as $FP$, False Negatives as $FN$, and True Negative as $TN$, The formula of sensitivity and specificity metrics \cite{lalkhen2008clinical} is as follows: 
\begin{eqnarray}
    \textrm{Sensitivity} &=& \frac{TP}{TP + FN}\\
    \textrm{Specificity} &=& \frac{TN}{TN + FP}
\end{eqnarray}

For computing true positives, false positives, false negatives, and true negatives, the IoU measure has been used (intersection over union) to determine the overlap between two boundaries divided by their union. Our dataset pre-defined an IoU threshold (i.e., 0.5) in classifying whether the prediction is a true positive or a false positive. Also, false negative would be those glomeruli objects that any predicted bounding boxes have not covered. Moreover, the true negatives were calculated based on the area of the whole slide tissue minus those predicted areas that were not containing any glomeruli.

As mentioned in the last section \ref{Chap:3}, 7 training datasets with two public and one private datasets have been created and validated on two datasets with different stains from the University of Michigan. In this section, the average sensitivity and specificity have been calculated for all these experiments for all  images, along with the comparison with the existing segmentation method.

\subsection{PAS Validation Set}
Two public datasets and a private dataset from the University of Michigan, all PAS stained, were used to train YOLO and validated on 20 PAS stained images. Different experiments were designed and evaluated on these images. Average sensitivity and specificity values for each experiment can be seen in Table  \ref{glomeruli-experiments-pas} along with the comparison with the segmentation method explained in Section \ref{Chap:3} (used for Hacking the Kidney Competition). 

\begin{figure*}[htb!]
\centering
%\captionsetup{justification=centering}
\centering
\includegraphics[height=8cm,keepaspectratio]{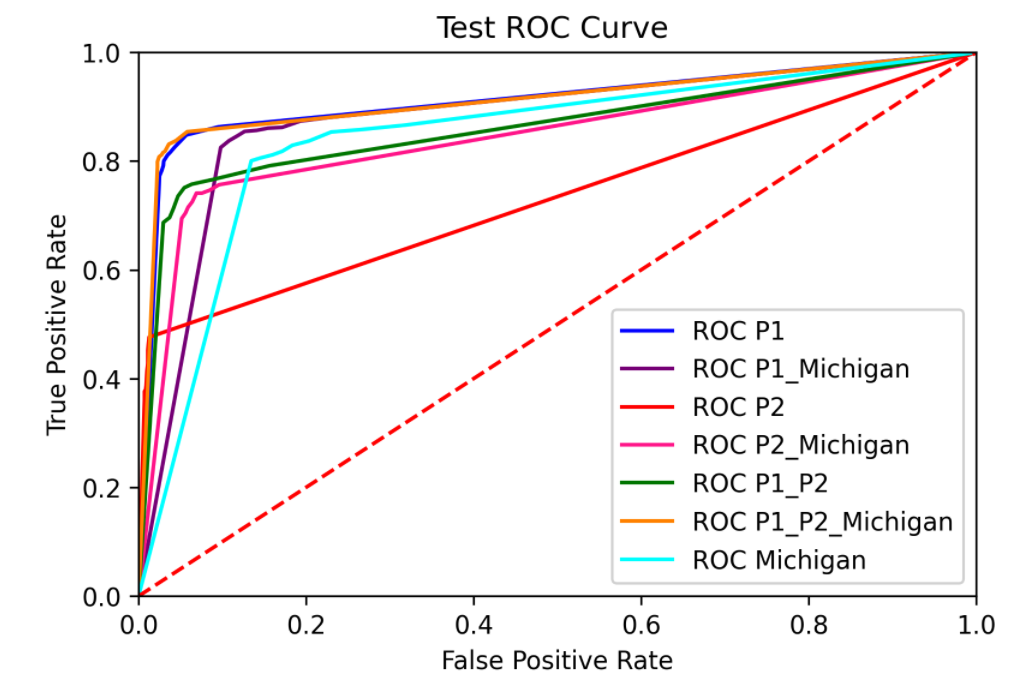}
\caption{ROC curve for 20 PAS stained images, and the comparison for all designed experiments using YOLO. $P1$ indicates the first public dataset, $P2$ indicates the second public dataset, and $Michigan$ is the data from University of Michigan for fine-tuning the models.}
\label{roc-pas}
\end{figure*}

The ROC (receiver operator characteristics) curves \cite{hoo2017roc} for all the experiments on these 20 PAS stained images have been shown in Figure \ref{roc-pas}. As it has been reported in Table \ref{glomeruli-experiments-pas}, the segmentation results have a high average specificity with lower sensitivity which means the network has low number of false positives. However, it can only predict half of the true negative glomeruli objects. Furthermore, using the external validation set (University of Michigan WSIs) to fine-tune this segmentation model is not feasible since the external WSIs do not contain pixel-level annotation.

Among the YOLO experiments, one experiment was done with a training set containing only 7 PAS stained images from the University of Michigan, with average sensitivity and specificity equal to 85\%, and 80\%, respectively, which can show the network is performing well on the validation from the same resource with limited training data. Another three experiments have been performed only on public datasets. They have been evaluated on an external validation dataset which is the data from the University of Michigan. 

By examining the network on an external dataset, the generalization of the network can be assessed. Also, it is evident that after fine-tuning the network with only 7 PAS stained images from the University of Michigan on the same dataset, the average sensitivity has a considerable improvement. For example, the average sensitivity and specificity changed from 45\%, and 98\% to 74\%, and 94\% respectively. The results may significantly change if there is more data of the same resource as the validation dataset for fine-tuning the network. 

Another important point would be the difference between the results of experiments trained on the first public dataset and the second one. It has been shown that by combining both datasets, the accuracy could drop off compared to only training on the first public dataset, and the reason may be related to the difference between the images from the second public dataset and the images from the University of Michigan. The images from the first public dataset and images from the University of Michigan are needle biopsy images. In contrast, the second public dataset consists of excision tissue samples. The phrase ``needle biopsy`` refers to a procedure in which a specific needle is inserted into a suspicious region of the skin in order to collect cells. During a ``surgical biopsy``, a surgeon creates an incision in your skin in order to reach the suspicious cells. As shown in Figures \ref{public_2_sample} and \ref{sample-michigan-wsi} number of glomeruli and the size of the glomeruli compared to the whole image are one of the differences between needle biopsy and surgical biopsy.

\begin{table*}[h!]
\begin{center}
\centering
%\captionsetup{justification=centering}
\caption{Average sensitivity and average specificity were reported for seven different experiments designed with two public datasets and a private dataset from UMICH (University of Michigan). All the PAS (Periodic acid–Schiff) stained images and evaluated on 20 PAS stained images, along with the comparison with a segmentation method using U-NET.}
\label{glomeruli-experiments-pas}
\resizebox{\linewidth}{!}{
\begin{tabular}{||c|c|c||} \hline
    \centering
    Dataset & Average Sensitivity & Average Specificity \\ \hline \hline
    \centering
    31 WSIs from public dataset 1 & 82\% & 95\% \\ \hline
    \centering
    31 WSIs from public dataset 1, fine-tuned with 7 PAS stained WSIs from UMICH dataset & 85\% & 89\% \\ \hline
    \centering
    8 WSIs from public dataset 2 & 45\% & 98\% \\ \hline
    \centering
    8 WSIs from public dataset 2, fine-tuned with 7 PAS stained WSIs from UMICH datase & 74\% & 94\% \\ \hline 
    \centering
    31 WSIs from public dataset 1, and 8 WSIs from public dataset 2 & 75\% & 95\% \\ \hline
    \centering
    31 WSIs from public dataset 1, and 8 WSIs from public dataset 2, fine-tuned with 7 PAS stained WSIs from UMICH dataset & 83\% & 96\% \\ \hline
    \centering
    7 PAS stained WSIs from UMICH dataset & 85\% & 80\% \\ \hline
    \centering
    Segmentation Method (HubMap Competition) & 48\% & 99\% \\ \hline
\end{tabular}}
\end{center}
\end{table*}

\subsection{H\&E Validation Set}

\begin{figure*}[htb!]
\centering
%\captionsetup{justification=centering}
\centering
\includegraphics[height=8cm,keepaspectratio]{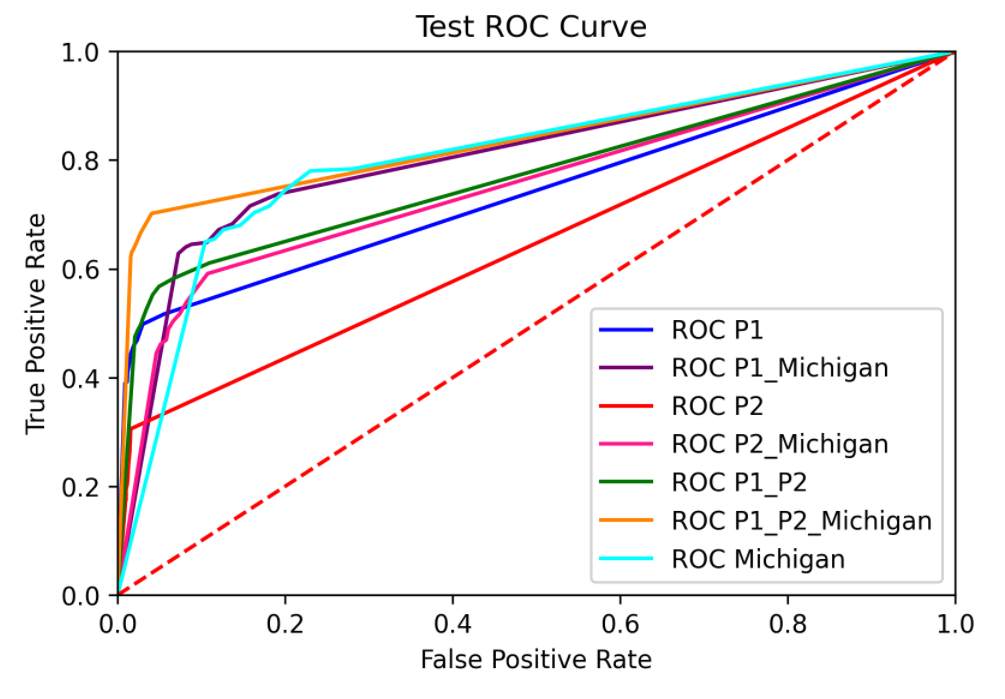}
\caption{ROC curve for 16 H\&E stained images, and the comparison for all the designed experiments using YOLO. $P1$ indicates the first public dataset, $P2$ indicates the second public dataset, and $Michigan$ is the data from University of Michigan for fine-tuning the models}
\label{roc-h-e}
\end{figure*}

A total of 16 H\&E stained images from the University of Michigan have been used as a validation dataset for all training datasets described in the previous section. Comparison between the average sensitivity and average specificity for all seven experiments using YOLO, with two public datasets, as well as a private dataset from the University of Michigan and the segmentation method explained in section  \ref{Chap:3} that was used for Hacking the Kidney Competition are provided in Table \ref{glomeruli-experiments-he}. ROC curves for all  experiments on these 16 H\&E stained images have been shown in Figure \ref{roc-h-e}. There is a considerable difference between the validation results on PAS stained images and H\&E stained images. This substantial difference is explainable because of the difference in tissue staining of training and validation datasets.

Same as in the Table \ref{glomeruli-experiments-pas}, because of the high average specificity and low sensitivity shown in Table \ref{glomeruli-experiments-he}, the network's segmentation results practically never show false positives. However, only half of the ground truth negative glomeruli objects can be predicted by this method. 

As well as the  Table \ref{glomeruli-experiments-pas}, the results have been improved by fine-tuning the training dataset with only seven images from the University of Michigan. The difference between the outcomes of experiments trained on the first public dataset and the second is still significant. Because of the differences in images between the second dataset which are surgical biopsy images and those from the University of Michigan that are needle biopsy images, it has been demonstrated that by combining both datasets, accuracy can drop. 

\begin{table*}[h!]
\begin{center}
\centering
%\captionsetup{justification=centering}
\centering
\caption{Average sensitivity, and average specificity reported for different seven experiments designed with two public datasets and a private dataset from  UMICH, all PAS stained and evaluated on 16 H\&E stained images, along with the comparison with a segmentation method using U-NET}
\label{glomeruli-experiments-he}
\resizebox{\linewidth}{!}{\begin{tabular}{||l|c|c||} \hline
\centering
    \centering
    Dataset & Average Sensitivity & Average Specificity \\ \hline \hline
    \centering
    31 WSIs from public dataset 1 & 51\% &  95\% \\ \hline
    \centering
    31 WSIs from public dataset 1, fine-tuned with 7 PAS stained WSIs from UMICH dataset & 67\% & 89\% \\ \hline
    \centering
    8 WSIs from public dataset 2 & 30\% & 85\% \\ \hline
    \centering
    8 WSIs from public dataset 2, fine-tuned with 7 PAS stained WSIs from UMICH datase & 59\% & 90\% \\ \hline 
    \centering
    31 WSIs from public dataset 1, and 8 WSIs from public dataset 2 & 58\% & 94\% \\ \hline
    \centering
    31 WSIs from public dataset 1, and 8 WSIs from public dataset 2, fine-tuned with 7 PAS stained WSIs from UMICH dataset & 70\% & 96\% \\ \hline
    \centering
    7 PAS stained WSIs from UMICH dataset & 70\% & 86\% \\ \hline
    \centering
    Segmentation Method (HubMap Competition) & 47\% & 99\% \\ \hline
\end{tabular}}
\end{center}
\end{table*}

\section{Conclusions}
There have been several technological advances across health care and digital pathology in recent years. Automated segmentation and pixel analysis of digital pathology images may identify diagnostic patterns and visual cues, leading to more reliable and consistent diagnostic categorization.

Glomeruli detection, as the first step of classifying the glomeruli diseases following by diagnosing different kidney diseases, is essential and critical in digital pathology. Because of the large number of these objects in the kidney, glomeruli detection could help pathologists save considerable time by computerized quantification. This paper trained YOLO-v4 with seven different training datasets consisting of two public datasets and a private dataset from the University of Michigan. Moreover, the networks were evaluated on 20 PAS stained images and 16 H\&E stained images from the University of Michigan. By training YOLO-v4 on the first public dataset, and fine-tuning by only 7 PAS stained images from the University of Michigan, experiments achieved 85\% average sensitivity and 89\% average specificity while validating the network on 20 PAS stained images from the University of Michigan, which was the best result out of different training datasets. For evaluating the network on H\&E stained images, 70\% average sensitivity and 96\% average specificity were obtained while training on both public datasets, followed by fine-tuning on the 7 PAS stained images. Also, final weights of a segmentation method based on U-Net have been used to evaluate the results on the same validation datasets. The model could achieve high specificity and lower sensitivity, making this method rather unreliable compared to YOLO with higher sensitivity. Moreover, obtaining pixel-level WSI annotations for the network is time-consuming. This makes the whole procedure for fine-tuning the model with limited data harder than detection methods like YOLO, which only requires a bounding box around the target objects.

\bibliographystyle{ieeetr}
\bibliography{references}
\end{document}